# Water Management Considerations for a Self-Sustaining Moonbase

Jeffrey S. Lee[1,2,3,4], Joe Yelderman[3], and Gerald B. Cleaver[1,4]

Early Universe Cosmology and Strings Group,
Center for Astrophysics, Space Physics, and Engineering Research[1]
Space Physics Numerical Modeling Group,
Center for Astrophysics, Space Physics, and Engineering Research[2]
Department of Geosciences[3]
Department of Physics[4]
Baylor University
One Bear Place
Waco, TX 76706

Corresponding author: Gerald_Cleaver@baylor.edu

**Abstract**

The most pragmatic first step in the all-but-inevitable 3rd-millennium Völkerwanderung of humanity throughout the Solar System is the establishment of a permanent human presence on the Moon. This research examines: 1. the human, agricultural, and technical water needs of a 100-person, 500 m × 100 m × 6 m self-sustaining lunar colony; 2. choosing a strategic location for the moonbase; 3. a heat drill model by which the needed lunar water ice could be sublimated; and 4. the robust water treatment and recovery infrastructure and water management personnel that would be needed for a self-sustaining moonbase.

## 1. Introduction

A self-sustaining moonbase is a lunar colony that is capable of the in-situ procurement of the three critical and immediate life needs – air, water, and energy.[1] Established in Section 2 are the three categories of water requirements for a self-sustaining moonbase: human needs, agricultural needs, and technical needs. For the entire moonbase complement, 12.3 acre·feet·yr$^{-1}$ of water will be sufficient for drinking, hygiene, shower, food preparation, dishwashing, and clothes washing. These essentials are independent of a person's location, and therefore, terrestrial human water needs form a blueprint for the human water needs on a moonbase. Additionally, oxygen for breathing and hydrogen for fuel and scientific studies can be easily obtained by water electrolysis.

The hydroponic farming of food to support the vegetarian diet of the moonbase personnel will require approximately 6 times the amount of water demanded by human needs (72 acre·feet·yr$^{-1}$). The water production requirements for the cooling of electronic equipment, firefighting, etc. are the least demanding, requiring only 2 acre·feet·yr$^{-1}$.

Section 3 considers the most suitable location for the moonbase. This site will require topographically flat regions that are consistently sunlit to provide solar power for the moonbase as well as consistently dark regions where frozen water can exist. This paper asserts that the most suitable moonbase location is along or adjacent to the Shackleton-de Gerlache Ridge at the lunar south pole.

---

[1] Clothing is not considered because it cannot be acquired in-situ.

The Lunar Prospector and the Lunar Reconnaissance Orbiter have provided conclusive spectroscopic data that demonstrate the presence of water ice and other condensable volatiles in this vicinity (Section 4).

The sublimation of sufficient lunar ice accompanying the adjoined lunar regolith can be accomplished with 6,524 10 kW heat drills. This is described by a 3D numerical model in Section 5. The sublimated material would be expelled from the lunar surface into a collection tent and then liquified in an adjacent liquefaction chamber. Following liquification, the water would be treated in the RFLUSH (a 5-stage water treatment facility) and made potable.

Section 6 proposes a militaristic or semi-militaristic administrative structure for the lunar colony. All water production, treatment, and distribution activities would be the responsibility of the Water Resources Manager who, as a member of the senior administration, would report directly to the Moonbase Commander. The Water Resources Manager would also be responsible for the required water provisions onboard the emergency evacuation spacecraft. Approximately 20% of the moonbase's 100-person contingent would be comprised of administrative personnel and specially trained technicians who are responsible for water sublimation, treatment, recovery, distribution, and management.

On Earth, water occurs as groundwater, surface water, and atmospheric water that is recycled and distributed in varying amounts among numerous environments. Water management strategies that are required on the Moon can offer insights into strategies for the management and sustainability of terrestrial water. Although the acquisition and management strategies for lunar water are still speculative, the necessity for improved water management on Earth is critically important for improved stewardship of a too often taken-for-granted but essential resource.

## 2. The Human, Agricultural, and Technical Water Needs

The smooth operation of the moonbase will be strongly dependent on the smooth operation of the water production, treatment, recovery, and distribution infrastructure and the professional expertise of its personnel.

Potable water would need to be continuously available for human activities and hydroponic farming. An additional small supply of water to support the electrolytic introduction and replenishment of oxygen will also need to be available.

### 2.1 Human Water Needs

The human water requirements on a permanent moonbase are expected to closely mimic human terrestrial water requirements because the human body's need for water are location independent. Showers, toilets, and washing machines use the greatest amount of indoor water [1]. Estimates for American household water needs vary, but 100 GPD·person$^{-1}$ is a typically accepted value [2]. This figure includes landscape irrigation, and as such, the human water needs on a moonbase would plausibly be less. However, for the purposes of this research, the 100 GPD·person$^{-1}$ figure is retained. Moonbase personnel will need to use their personal water allotment for the same indoor purposes as terrestrial families. Consequently, the distribution in Figure 1 is likely to be applicable to the lunar environment.

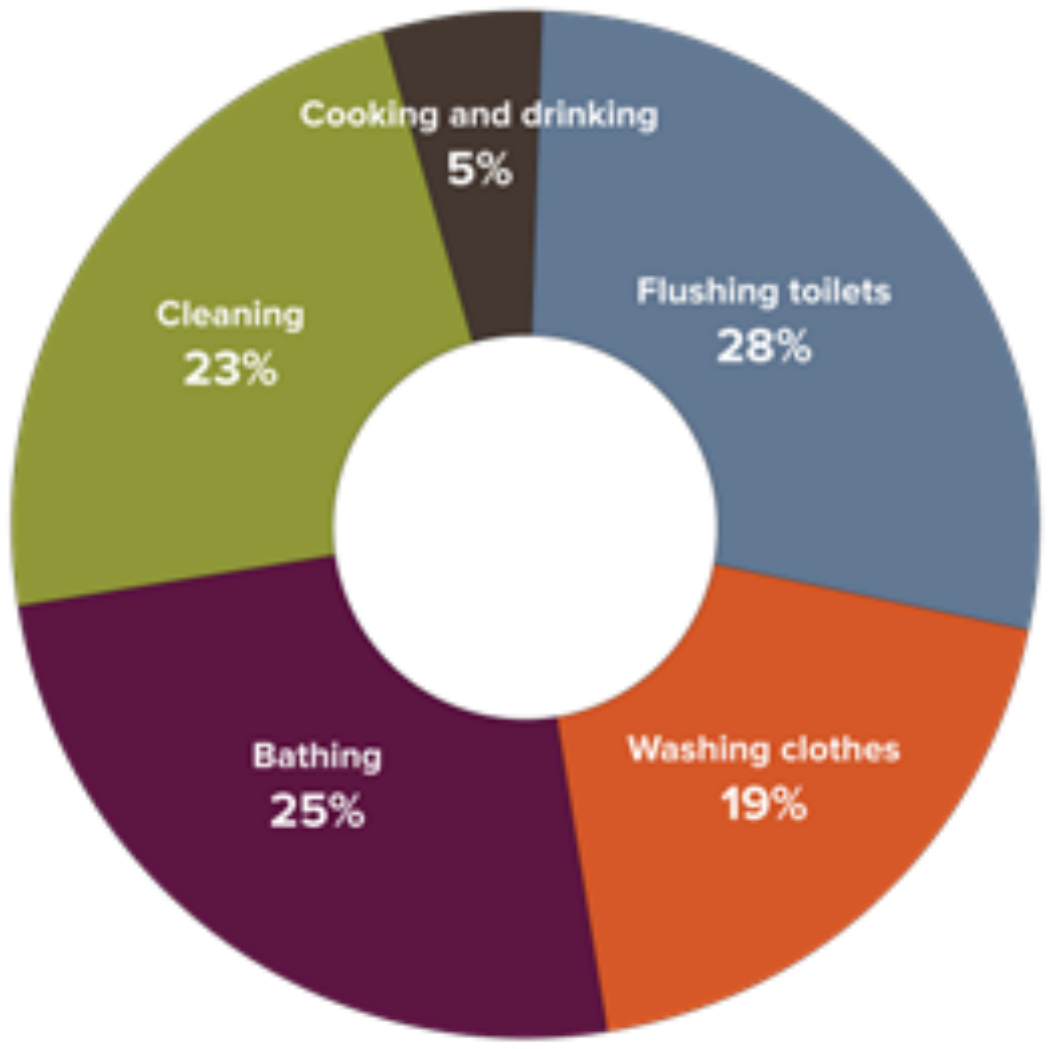

Figure 1: Household indoor water use in the United States (2016) [3].

For a 100-person lunar colony, 100 GPD·person$^{-1}$ with a 10% buffer results in a total requirement for personal water use of 11,000 GPD or 12.3 acre·feet·yr$^{-1}$.

### 2.1.1 **Electrolytic Production of Oxygen**

The terrestrial atmosphere is composed of 21% oxygen and 78% nitrogen with trace amounts of argon, carbon dioxide, helium, neon, methane, krypton, and hydrogen. A moonbase atmosphere would need to be similar (21% $O_2$ and 79% $N_2$ would suffice); the trace gases are not essential for human wellbeing. Both green plants and carbon dioxide scrubbers would be put in place for the removal of exhaled $CO_2$. However, the required 78 tons of $O_2$ for breathing (see eq. (1)) would need to be either transported from Earth (in addition to the 273 tons of $N_2$)[2] or obtained by the electrolytic separation of water.

---

[2] The $N_2$ mass calculation is the same as the $O_2$ mass calculation, except the relative partial pressure of $N_2$ is 0.79.

$$n = \frac{PV}{RT} = \frac{\left(1.01325\times10^{5}\ \text{Pa}\right)\left(0.21\right)\left(500\ \text{m}\right)\left(100\ \text{m}\right)\left(6\ \text{m}\right)}{\left(8.314\ \text{J}\cdot\text{mol}\cdot\text{K}^{-1}\right)\left(276\ \text{K}\right)} \rightarrow m = 78\ \text{tons} \qquad (1)$$

At the time the moonbase construction is completed and the colony is ready for occupancy, the total volume of water that would need to be electrolytically converted to $O_2$ for breathing is 13,000 gallons (0.04 acre·feet).

Whether oxygen would be obtained electrolytically from water and/or from exhaled $CO_2$, or whether it would be brought from Earth by resupply missions is an engineering and design decision that is beyond the scope of this paper. However, it seems likely that all three methods should be considered because oxygen is the single most important quantity for the survival of the moonbase inhabitants.

Fortunately, however, the water requirements for the complete electrolytic replenishment of the breathable oxygen are only 0.3% of the personal water requirements for the entire moonbase's complement.

The hydrogen produced by the electrolysis of water would also be extremely valuable as fuel and for laboratory experiments. Specialized equipment, such as Sunfire's Pressurized Alkaline Electrolyzer (SPAE) (see Figure 2) could produce hydrogen independently. Although not investigated here, the need for hydrogen is unlikely to significantly affect the water budget.

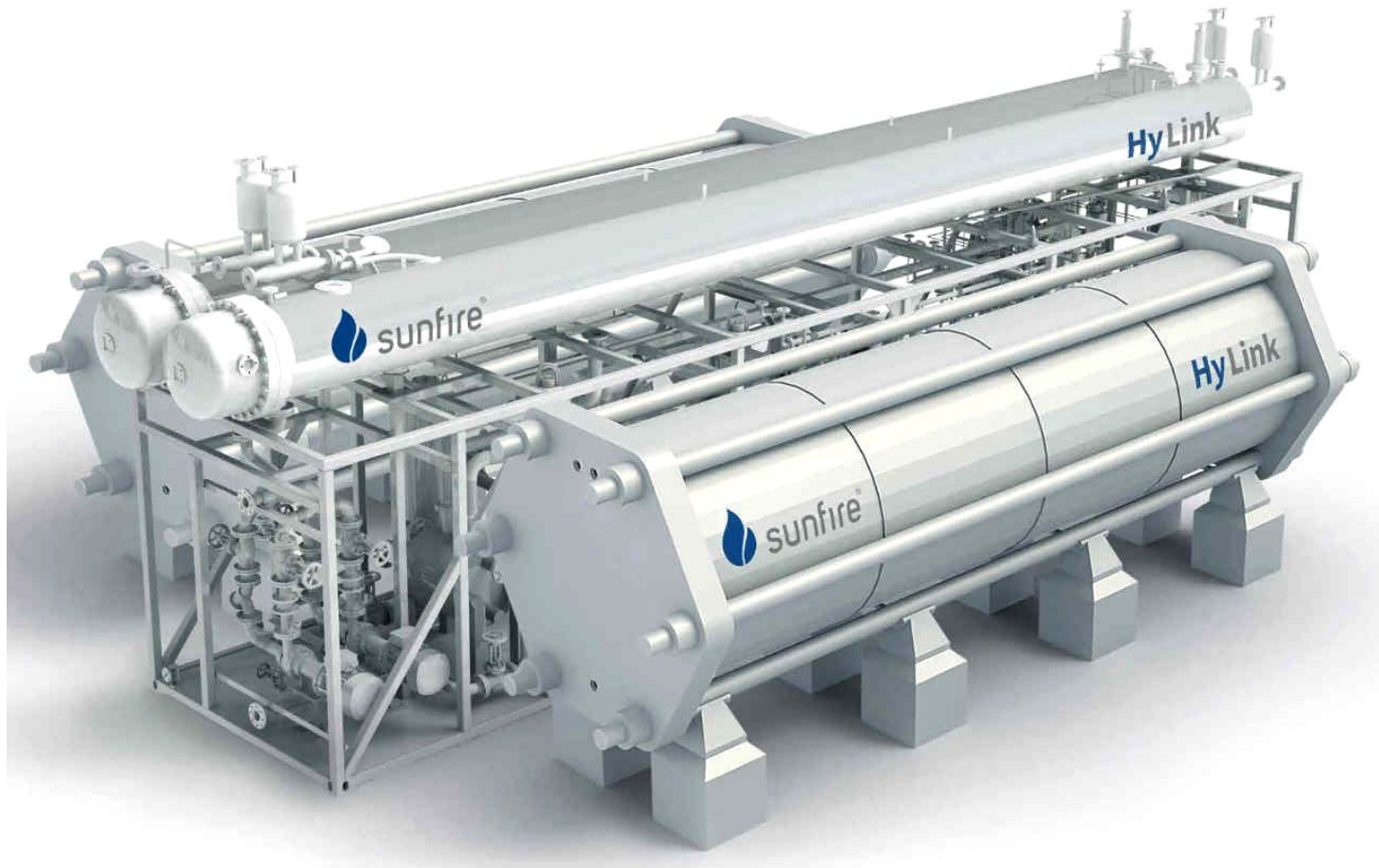

Figure 2: Sunfire's Pressurized Alkaline Electrolyzer capable of producing 2,150 $Nm^3 \cdot hr^{-1}$ of hydrogen at 30 bar and consuming 4.7 $kWh \cdot Nm^{-3}$.[3] Its scalable design allows the concatenation of numerous SPAE units.

## 2.2 Agricultural Water Needs

The 7-8% energy conversion efficiency of cattle [4] makes strictly vegetarian diets essential on the moonbase, and no food production animals would be in residence.[4]
The lunar environment makes enclosed hydroponic farming the only viable means for long-term food production. The required farming area is estimated to be 370 $m^2 \cdot person^{-1}$ [5].
At 6 $L \cdot m^{-2} \cdot day^{-1}$ (1.6 $GPD \cdot m^{-2}$), the moonbase personnel would require 72 $acre \cdot feet \cdot yr^{-1}$ (23.5 million $gallons \cdot yr^{-1}$) to grow the food necessary for their sustenance.

Although hydroponic plant growth in lunar regolith [6-12] is better understood than it is in lunar gravity (0.167*g*) [13-18], the general consensus in the literature supports the assertion that lunar hydroponic farming is possible and feasible. However, prior to the approval of the moonbase's construction, successful, small-scale, and in-situ hydroponic experiments would need to be conducted to confirm the terrestrial and low Earth orbit laboratory results.

[3] $Nm^3$ refers to "normal meters cubed" (the volume at $0^o$ C and 1 standard atmosphere).

[4] It is, of course, plausible that special occasion, non-vegetarian meals could be delivered to the moonbase on resupply shuttles.

**2.3 Technical Water Needs**

The amount of water required to support the technical activities on the moonbase is more speculative. Cold water would be needed to cool Aquasar™ supercomputers, "supra-Schwinger" lasers[5], interstellar cubesat propulsion lasers[6], as well as for general scientific experiments.

Additionally, fire suppression stations throughout the moonbase would need to be established and connected directly to specially designated water storage tanks. Combustion in reduced gravity environments [19-21] is poorly understood, and firefighting in lunar gravity even less so.

The watertight hatches design by which a modern naval vessel is internally compartmentalized could be modified into airtight hatches which section off the interior of a moonbase. Incorporated into this design would be an airlock in each section that could be individually and remotely operated. If a fire were to occur, the affected section could be evacuated, and the airlock could be opened, exposing that section to the vacuum of space, and denying the fire a supply of oxygen. This technique is extreme because in addition to the need to repressurize the evacuated section after the fire is extinguished and the airlock hatch is closed, all items not secured in place will be violently hurled into space as the local atmosphere is supersonically expelled. Furthermore, security precautions would need to be implemented to prevent the enactment of a nefarious plan by a diabolical individual who is intent on opening the airlock hatch of a moonbase section that is occupied by unprotected personnel. It is likely that this fire suppression procedure would be used only as a last resort, and therefore, this contingency does not significantly affect the water budget of the lunar colony.

---

[5] Lasers with beam intensities greater than the Schwinger Limit ($10^{29}$ W/cm$^2$) (the scale above which the electromagnetic field ceases linearity).

[6] High intensity lasers supplying the photon pressure to accelerate tiny cube-shaped satellites to relativistic speeds (~0.2*c*) for interstellar voyages.

It is also not clear that water would always be the ideal firefighting medium. It may be that PFAS foams are more efficacious against some fires in the moonbase, although by no means has this been demonstrated for a lunar colony. Regardless, it is prudent that the specifications of lunar firefighting stations are at least equal to the minimum requirements of their terrestrial counterparts – 500 gpm and 20 psi [22].

To this end, the figure of 2 acre·feet·$yr^{-1}$ (650,000 gallons·$yr^{-1}$) is selected and is predicated upon the realistic assumption that the moonbase energy will come from the likely combination of solar power and $^{87}Rb$ β decay. Neither of these power sources require extensive cooling measures. However, were the moonbase to be powered by a fission reactor, the water needs would increase substantially.

Fission reactor cooling requires from 400 to 720 gallons·$MWhr^{-1}$ of cold water [23]. Thus, a large nuclear power plant could require 1.1 million acre·feet·$yr^{-1}$ (360 billion gallons·$yr^{-1}$). Consequently, from the water requirements alone, lunar nuclear fission power plants appear to be infeasible.

## 2.4 Total Water Needs

Human, agricultural, and technical water requirements total 86.3 acre·feet·$yr^{-1}$ (neglecting the nuclear fission option) and are compared in Figure 3. If the allotted 2 acre·feet·$yr^{-1}$ is either excessive or insufficient, it represents only 2% of the total water budget and is therefore not highly impactful. Agricultural needs represent the largest water requirement (84% of the total water budget). This is fortunate because were the food-producing ecosystem to fail utterly and irreparably, a total evacuation of the moonbase ("Operation Exodus", see Section 6.3.1) would need to commence. The personnel could return to Earth after a 3–4-day voyage, and the complete collapse of the water-supporting biome would not endanger the inhabitants.

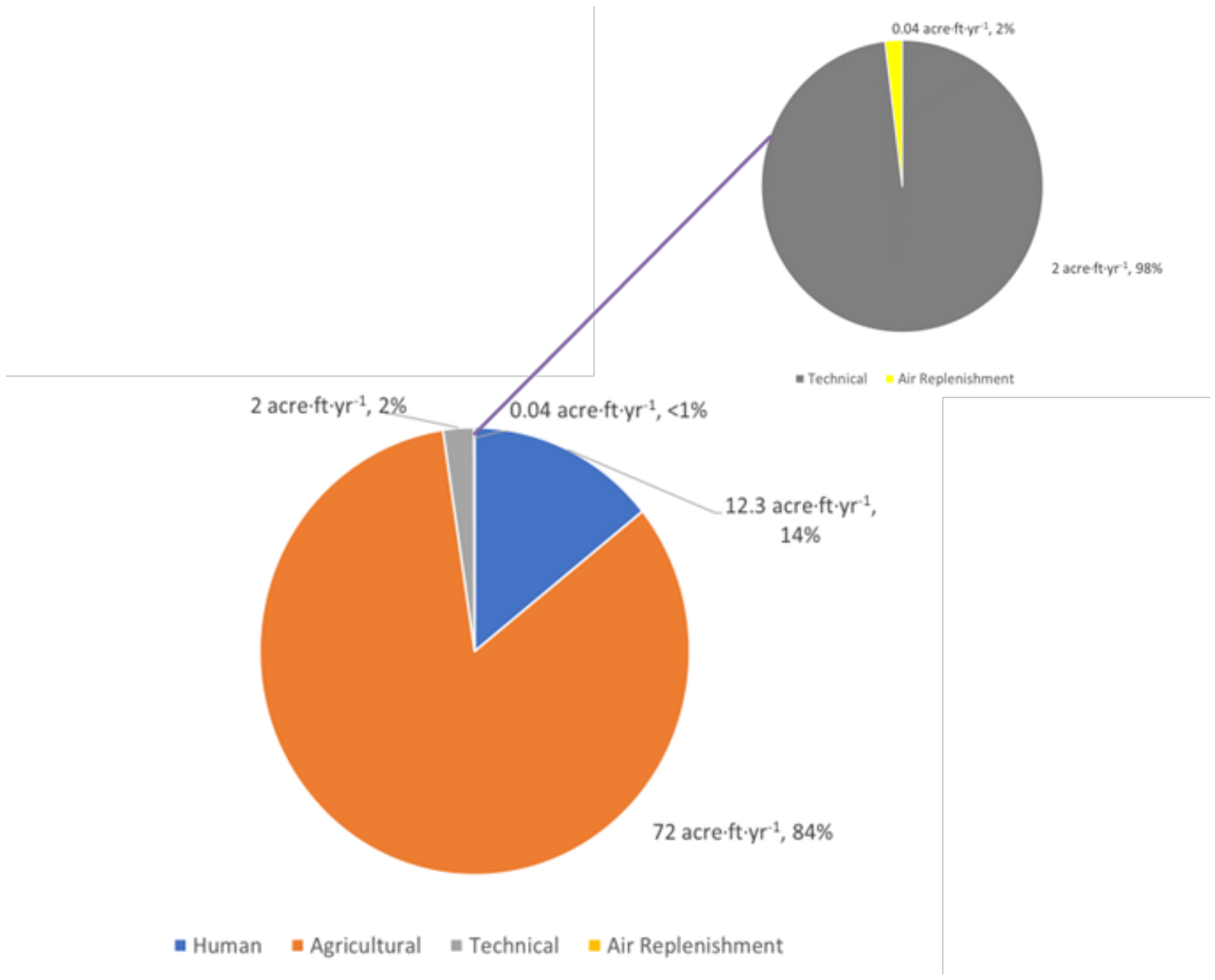

Figure 3: Comparison of the human, agricultural, and technical water needs of a self-sustaining moonbase.

## 3. Selecting the Moonbase Location

Significant real estate is available on the Moon. The lunar surface area is $3.793 \times 10^7$ km$^2$ which is the combined area of the United States, Canada, and Russia. Selecting a suitable location for the lunar colony must consider a number of factors. For instance, the terrain must be topographically flat. The base needs to be in or near a sunlit region which is proximate to the shadowed plateau within an impact crater where lunar ice would be abundant to supply the needed water. Water needs force the moonbase location to be at one of the lunar poles, where infrared reflectance spectroscopy conducted by the Chandrayaan-1 spacecraft has revealed the characteristic absorption peaks of $6\times10^{11}$ kg (486,000 acre·feet) of water ice [24]. As shown in Figure 4, the ice is located predominantly in the south polar region.

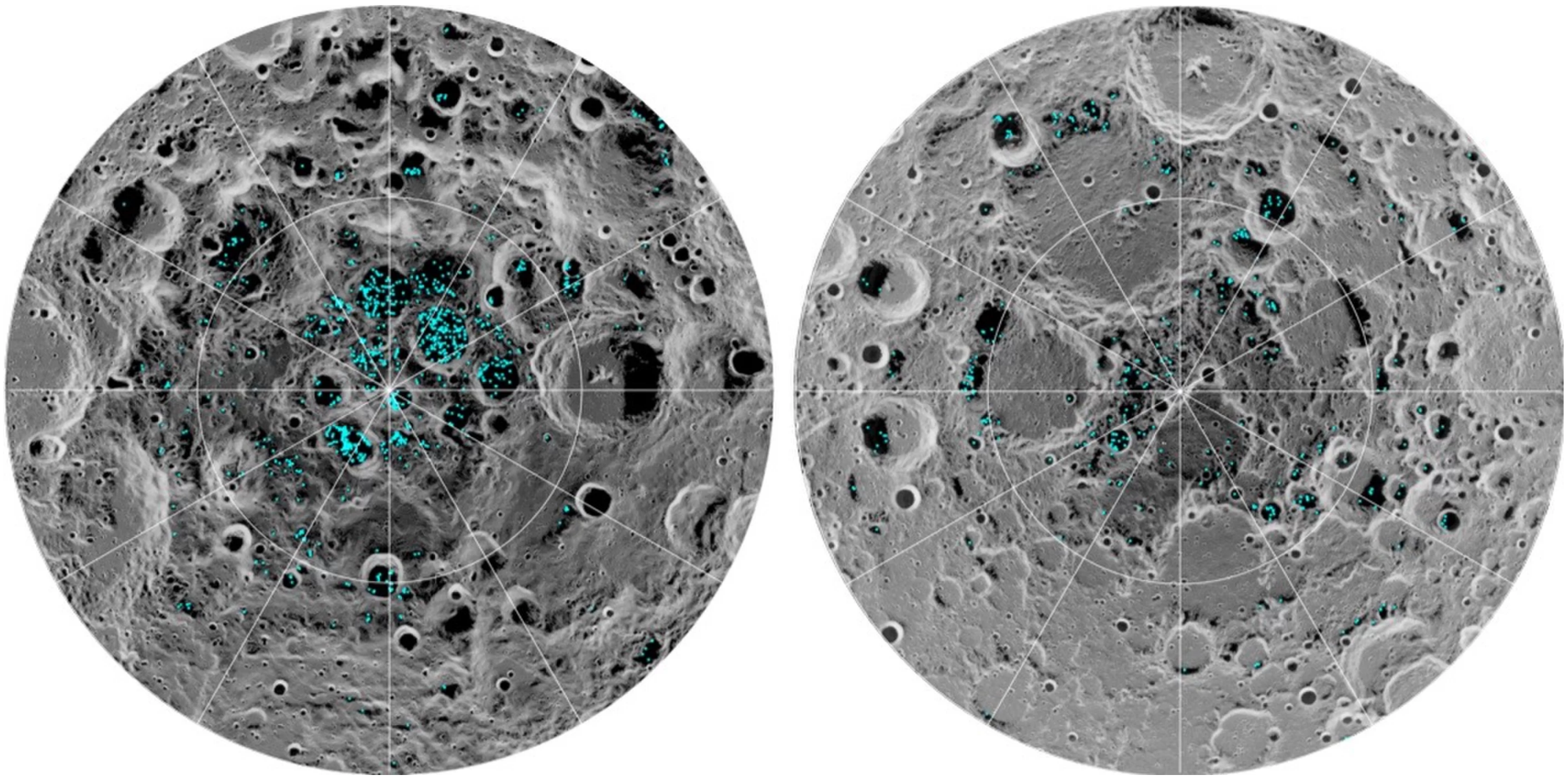

Figure 4: Locations of lunar ice (indicated in turquoise) at the south pole (left) and north pole (right) [25].

Perhaps the most promising location for the moonbase is on, or adjacent to, the Shackleton-de Gerlache Ridge located at 89.9ºS 0.0ºE (see Figure 5, Figure 6, and Figure 7). The Ridge luminosity is ideal because three rim locations are sunlit 90% of the time, and the region is circumscribed by always-dark topographic depressions, where the permanently shadowed regions retain temperatures of 23 K, and lunar water ice is abundant.

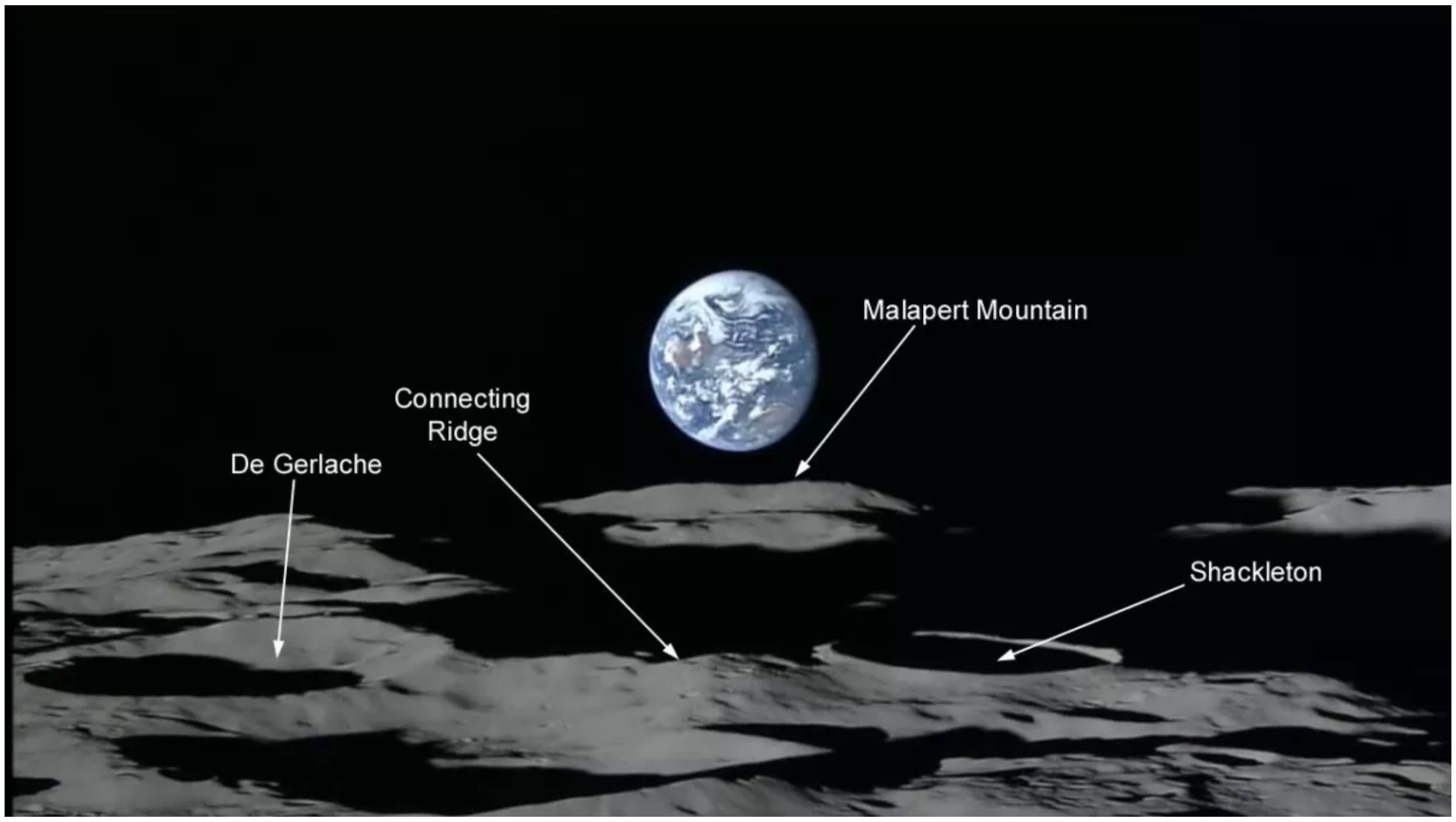

Figure 5: The Shackleton-de Gerlache Ridge with the Earth overhead [26].

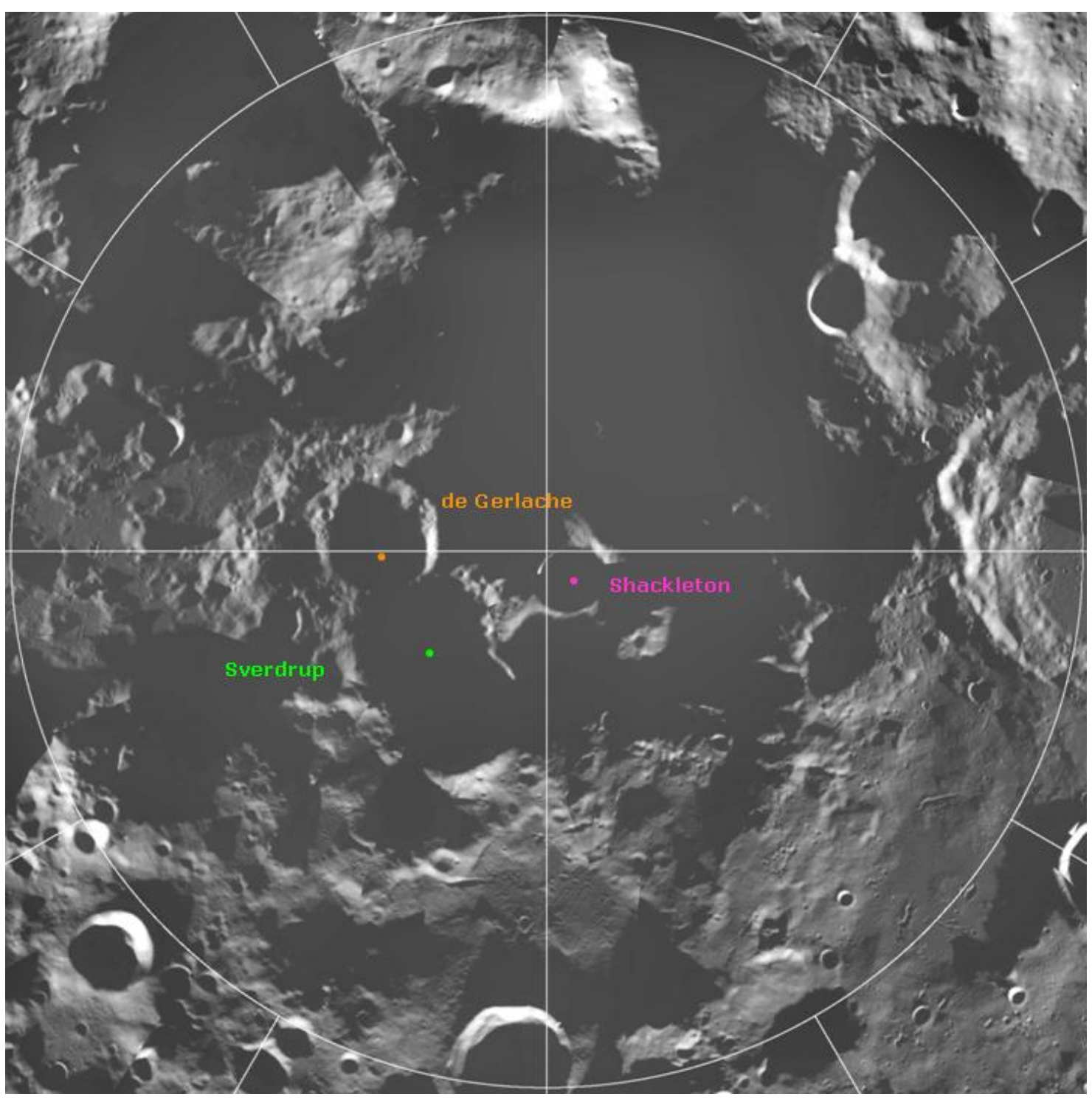

Figure 6: The Shackleton and de Gerlache craters as imaged by the Clementine spacecraft [27].

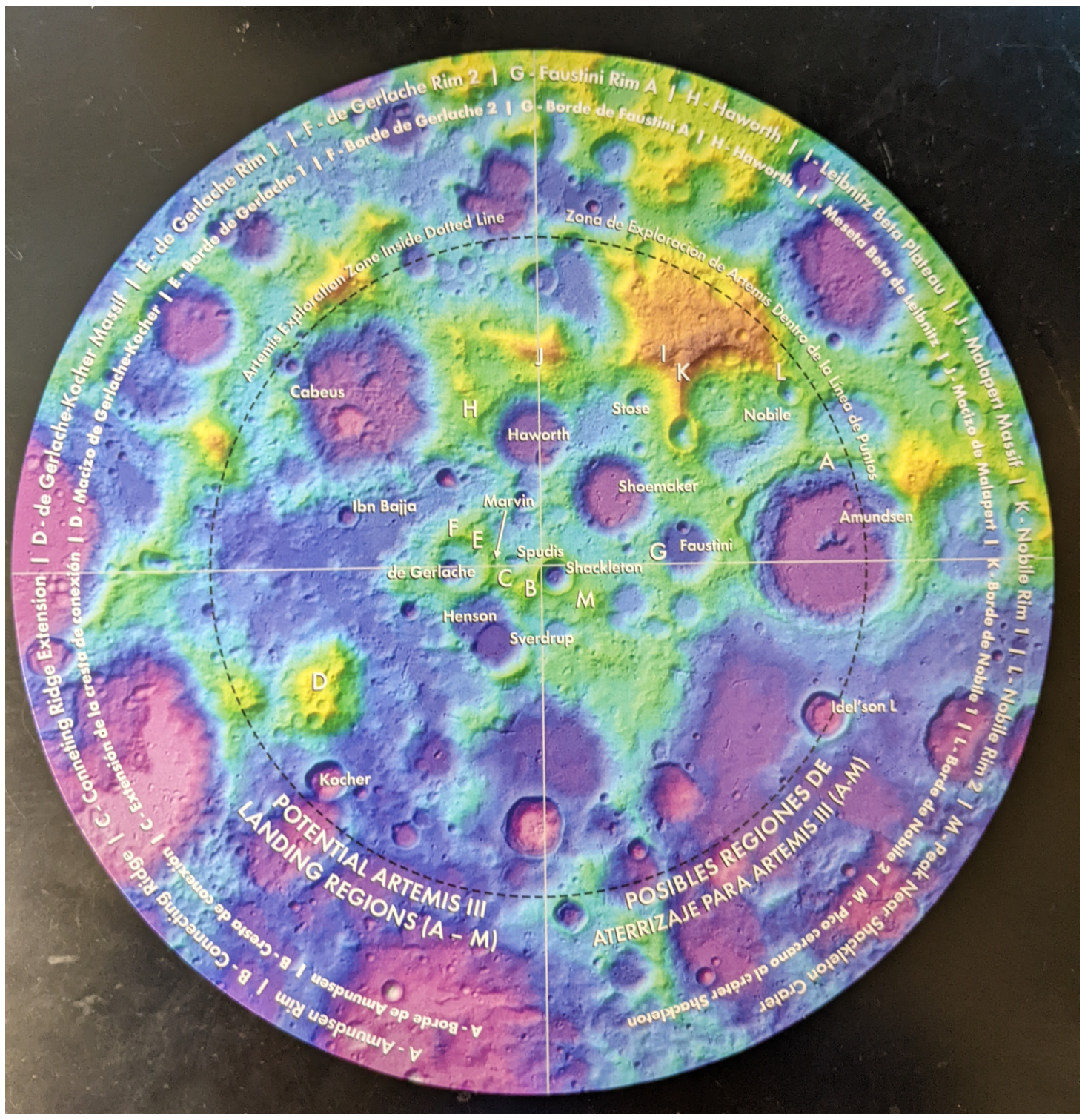

Figure 7: Topographical map of the southern polar expanse indicating possible landing regions for the (human return to the Moon) Artemis III space missions [28]. Pink and brown colorations indicate the regions of lowest and highest elevation, respectively.

## 4. The Credible Evidence for Lunar Water Ice

The presence of a permanent, self-sustaining moonbase is entirely dependent on the ability of such a moonbase to generate its own life essentials (useable energy, breathable air, and potable water). Were the Moon to lack an abundant water supply, a moonbase would be rendered impractical, if not altogether impossible. At 1,000 $kg \cdot m^{-3}$, the annual water needs of the lunar colony would require the annual Earth-to-Moon transportation of $10^8$ kg of water. Even with ambitious water recycling which could reduce the water needs by up to 97% (discussed in Section 6), more than $10^6$ $kg \cdot yr^{-1}$ of water would nevertheless need to be brought from Earth. Consequently, in-situ water acquisition is essential.

Lunar In-Situ Resource Utilization (ISRU) has been discussed extensively in the literature (e.g., [29-38]) due to its requirement for future lunar commercial and exploration endeavors. Abundant concentrations of condensable volatiles, including water ice, have been credibly observed at the lunar poles. Lunar Prospector has measured epithermal neutron fluxes emerging from the lunar regolith [39] as has the Lunar Exploration Neutron Detector onboard the Lunar Reconnaissance Orbiter (LRO) [40].

Studies of water transport across the lunar surface have indicated that 20-50% of juvenile water is likely retained in a frozen state within the polar lunar craters [41-45]. Although this was initially contested because interstellar hydrogen Lyman-α dissociation [46], particle sputtering-induced erosion [47], and losses resulting from meteoric bombardment [41] may be collectively in excess of the accretion rate, thus preventing permanent polar ice deposits, the evidence for lunar water ice remains stalwart.

Simulated neutron flux spectra (neutron lethargy as a function of neutron energy of $H_2O$-laden ferroan-anorthosite (FAN)[7]) indicate that hydrogen, in the form of water molecules, monotonically decreases epithermal and fast neutron intensities [39]. This result is borne out by the Lunar Prospector data of the period from January 16 – June 27, 1998 in which thermal and epithermal counting rates were formed from the $^3$He(Cd) and $^3$He(Sn) pulse-height spectra sums. These results are consistent with 1850 km$^2$ deposits of water ice beneath 40 cm of dry regolith at the lunar poles.

---

[7] A significant constituent of lunar highland regolith.

The Diviner Radiometer onboard the LRO has revealed that within the polar craters, as well as less than 1 m below an overburden in the surrounding regions, water stability is present, and lunar ice sublimation will not naturally occur [48]. Importantly, Gladstone et al. [49] have shown that water-frost-consistent ultraviolet emissions, detected by the Lyman Alpha Mapping Project instrument onboard the LRO, are suggestive of abundant subsurface water resources laden in the lunar regolith.

Although the presence of water ice on the Moon is a highly credible assertion, its origin remains controversial. One of the most accepted theories is the cometary and asteroidal delivery of copious quantities of water to the lunar surface during the Late Heavy Bombardment (4 Ga) and continuing to as late as 1 Ga. The impact-ejected water molecules formed a temporary atmosphere, most of which escaped the low lunar gravity. However, the remaining water descended and became frozen to the lunar floor cold traps within the shadowed regions of impact craters. Regardless of the origin of the lunar ice, ambitious but feasible sublimation schemes have been proposed for its recovery.

## 5. The Sublimation of Lunar Water Ice: A 3D Numerical Model

Water extraction from the lunar regolith has been studied by Biswas et al. [50] and Sowers and Dreyer [33]. Among the most promising approaches, however, is a 3D numerical model created by Brisset et al. [51] in which impermeable and homogeneous iron rods are quadratically distributed on the surface of the Moon and inserted to a depth of 50 cm. Each heat drill services a regolith volume of $1.2 \times 1.2\ m^2 \times 1$ m (depth) (see Figure 8). The rods are heated with constant powers ranging from $1–10^4$ W.

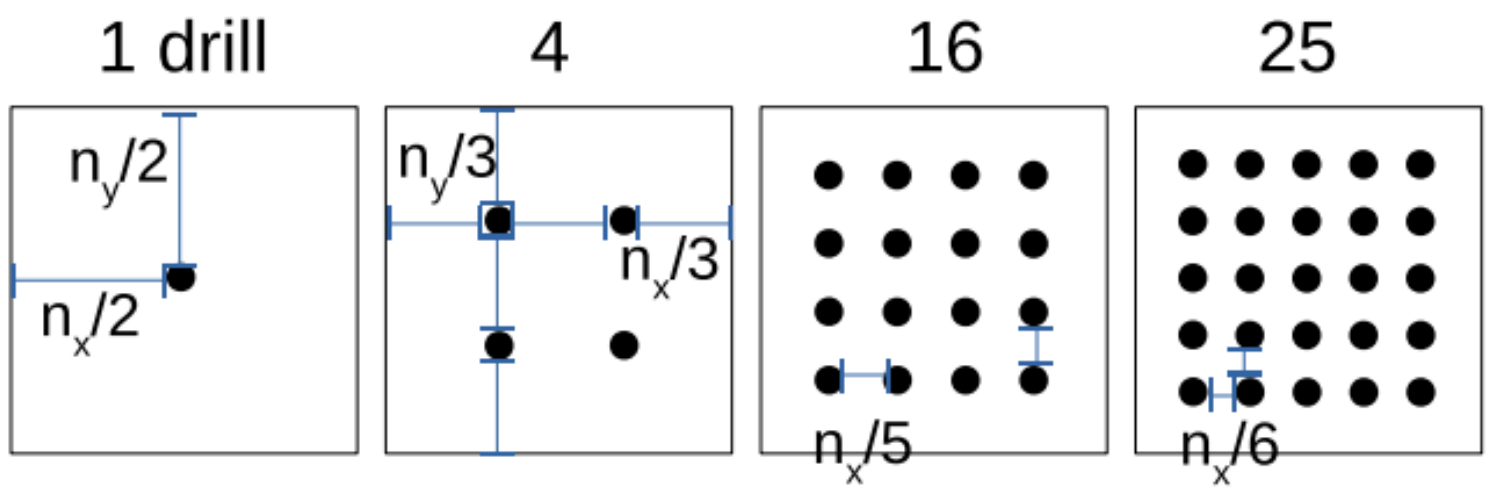

Figure 8: Quadratic drill distribution shown in a 2D lattice [51].

The model accounted for thermal diffusion in which radiative heat transfer between regolith grains was neglected. Turbulent flow is assumed to be absent[8], and Darcy's Law is applied along each cell dimension, from which the vapor masses undergoing intercellular transport are calculated. The Arden-Buck equations [52] produced an ice-vapor saturation profile (see Figure 9) in which the von Neumann boundary condition $\nabla T = 0$ was applied to the sides and bottom of the total 3D lattice. This ice-vapor saturation profile is expected because the vapor pressure is (and should be) dependent only on the temperature [53, 54].

The near-one-terrestrial-atmosphere ice-vapor saturation pressure achieved when the regolith temperature reaches 100$^{o}$C strongly indicates the copious outgassing of sublimated volatiles from the lunar surface.

[8] This is a reasonable assumption because intercellular explosive outgassing of condensable volatiles is not expected due to their relatively low abundance. As such, the Forchheimer corrections to Darcy's Law do not need to be invoked.

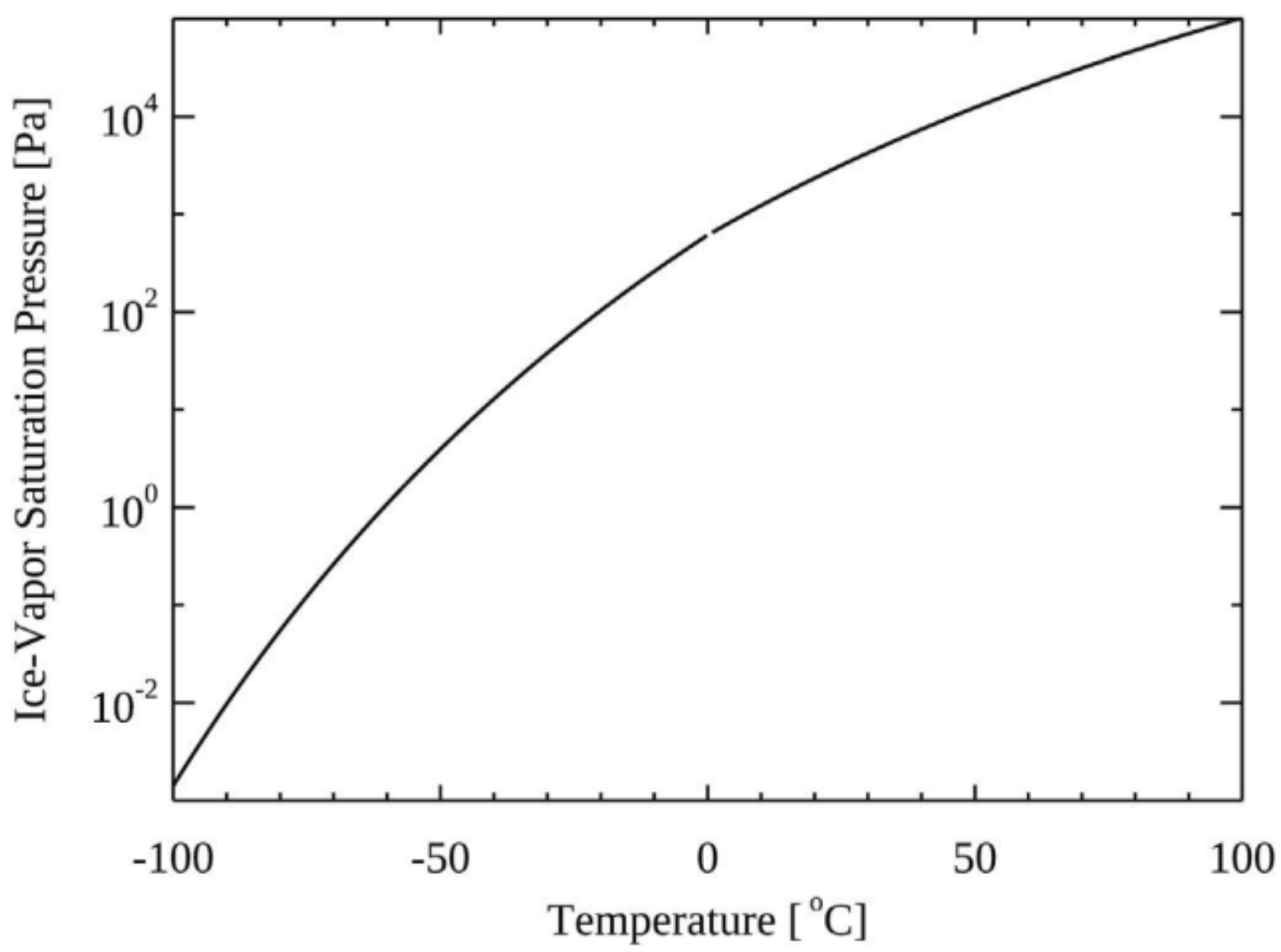

Figure 9: Water ice-vapor saturation vapor pressure from the Brisset model and determined with the Arden-Buck equations [51].

The Dirichlet boundary condition for the regolith temperature at the surface of a heated volume $(T = const.)$ is determined by the power applied to the associated heat drill. The lunar surface atmospheric pressure of $3\times10^{-10}$ Pa [55] is easily achievable in practice. The drill-regolith interface constitutes a third heating boundary, and the Dirichlet boundary condition at this interface is also governed by the heat drill power. The impermeability of the heat drills ensures that all vapor flux is either intercellular or is emitted from the surface.

The results of the Brisset model (shown in Figure 10) indicate that for the case of 1% ice abundance in the regolith, 1 kg of lunar ice can be sublimated in ~100 min. To achieve the total water requirements of the moonbase (86.3 acre·feet·$yr^{-1}$), 20,240 10 kW heat drills are required to operate continuously across a 170 × 170 $m^2$ regolith sublimation field.

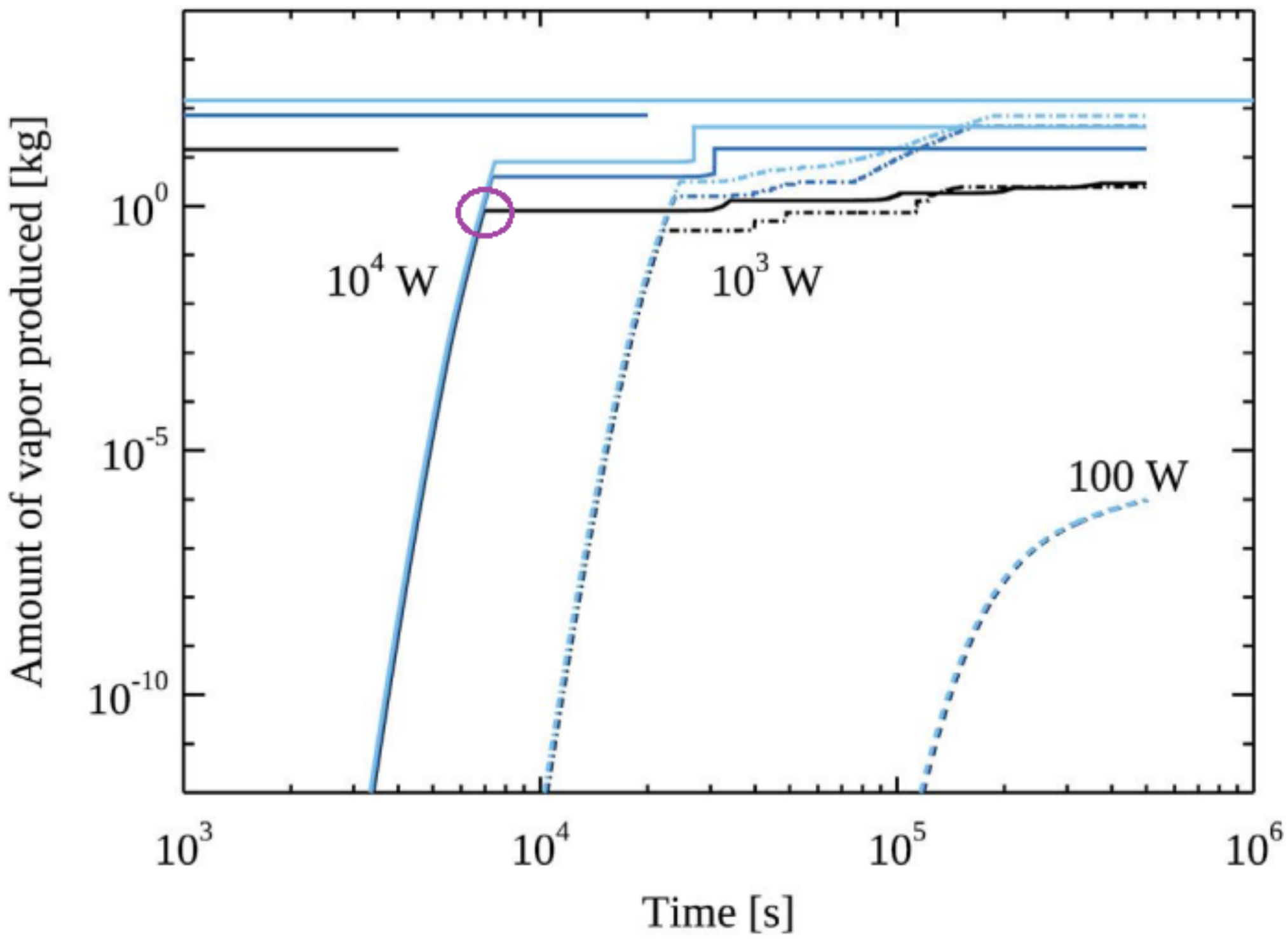

Figure 10: Mass of vapor produced from surface heating. The dotted and dashed lines indicate the power applied to the heat drills. The black curve represents an ice abundance of 1% in the lunar regolith; similarly, the dark blue curve indicates a 5% abundance, and the light blue curve indicates a 10% abundance. The purple circle represents the most conservative estimate of ice-vapor production (1 kg per ~6,000 s = 0.01 kg·min$^{-1}$).

### 5.1 Recovery of Sublimated Water Ice Vapor

The $3\times10^{-10}$ Pa lunar surface pressure and the 1.63 m/s$^2$ acceleration due to gravity guarantee that the sublimated lunar regolith and ice-vapor are ejected with sufficient speed so as to not immediately fall back upon the lunar surface.

Situating the heat drills within a $3\times10^{-10}$ Pa capture tent (see Figure 11) from which the ice vapor can be extracted through exhaust vents into a connected $3\times10^{-10}$ Pa liquefaction chamber provides a method to recover the sublimated water ice.

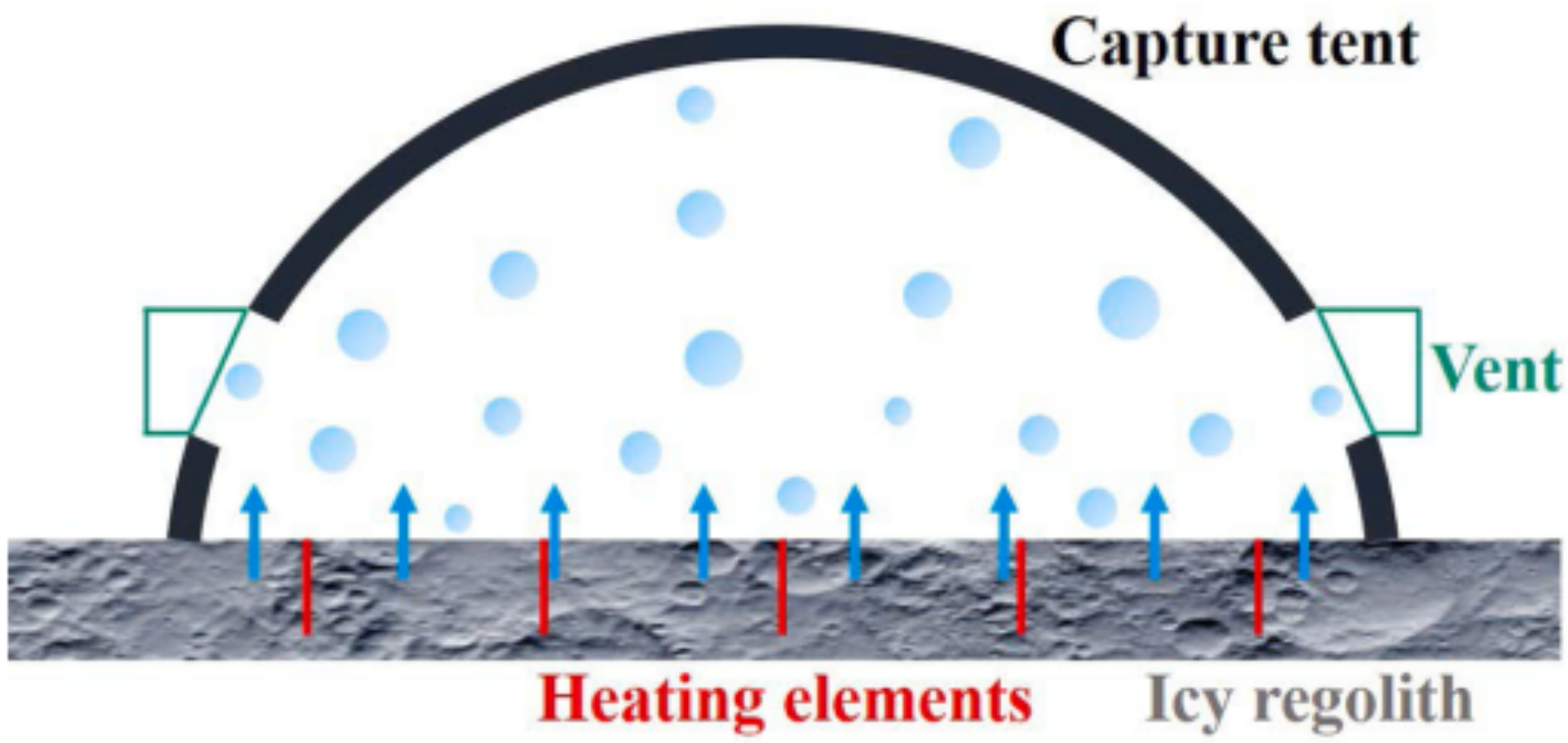

Figure 11: A proposed capture tent for sublimated lunar water ice. The ambient lunar atmosphere partial pressure within the capture tent would be kept at ~$3\times10^{-10}$ Pa [56].

Although the saturation pressure of ice-water vapor (~$10^5$ Pa) is four orders of magnitude greater than the estimated vapor partial pressure of the volatiles in the capture tent (~18 Pa at 240 K [56] because the capture tent is not vapor saturated), there will nonetheless be a negative pressure condition. Thus, the sublimated material is swiftly expelled from the 18 Pa capture tent into the $3\times10^{-10}$ Pa liquefaction chamber (not shown in Figure 11). The optimal temperature[9] within the liquefaction chamber is uncertain and is a subject for future research. If the liquefaction rate is substantially less than the sublimation rate, the moonbase's water needs would not be achieved by the estimated 20,240 10 kW heat drills, and a more substantial water recovery architecture would be required. The investigation of such an architecture is beyond the scope of this paper.

[9] The temperature that maximizes the rate of liquefaction.

## 6. Water Treatment and Recovery

The infrastructure that would support the water needs of a self-sustaining lunar colony is vastly in excess of that of even long-duration low-Earth-orbit spaceflight (e.g., 2 years on the International Space Station). The water recovery systems onboard modern spacecraft employ oxidizing pretreatment chemicals because filtration or distillation systems, even those designed for a microgravity environment, are subject to fouling and even irreparable failure. For water recovery on the moonbase, well-established terrestrial technologies (solar disinfection, distillation, and media filters) could be used. Instead of wastewater (e.g., urine) stabilization in preparation for disposal, the wastewater could be permitted to create filterable precipitates and biofilms within the treatment media. That water could then be filtered and recovered for future use.

For the design of terrestrial water treatment facilities, the principal drivers are the quality of the influent water and the desired quality of the effluent water. Reliability, cost, and location-based constraints also factor into the design. Physical, chemical, and biological treatments are necessary, but each is effective only against certain classes of contaminants. No single treatment can transform the influent wastewater into potable water. A multiple barrier treatment train is an advisable strategy, particularly when pathogen contaminants are a concern; furthermore, its intrinsic redundancy provides a reliability increase [57].

These considerations will all apply to a moonbase's water treatment facility, and as such, this facility is likely to closely mimic its terrestrial counterparts.

The aggregation of colloidal particles into flocs, accelerated by the addition of chemical coagulants, which are removed by settling under the influence of gravity, is a time-tested technique for terrestrial water treatment. Although successful flocculation on a moonbase is reasonably expected, additional modeling of the lunar gravity settling of flocs should be undertaken. On Earth, 90% of the required bacterial decontamination can be provided by the flocculation stages [57].

Filtration will also be essential, and there are four categories of membrane filtration systems in terrestrial use: reverse osmosis, nanofiltration, microfiltration, and ultrafiltration. These four categories have different operating pressures, constituent rejections, and pore dimensions. Microfiltration and ultrafiltration rely on size exclusion, whereas the more selective reverse osmosis and nanofiltration rely on osmotic events [57].

Gravity-fed granular filters could, in principle, operate even in lunar gravity because direct pressurization is also an option. Without direct pressurization, it is uncertain whether these filters could operate in 1/6 of Earth's gravity. Even in lunar gravity, rapid sand filters will probably remove coagulant-treated contaminants at a rate greater than terrestrial slow sand filters which are 100 times slower than terrestrial rapid sand filters. It is conceivable that lunar rapid sand filters could form a very thin schmutzdecke to aid in the filtration of influent contaminants, although this would almost certainly be a weak, if not negligible, effect. The use of sand filters to remove bacteria and particulates, as well as granular filters containing anthracite coal could be used in the water treatment facility of a moonbase. However, lunar regolith is an alternative in-situ filter media which is available in near-unlimited abundance.

It is unclear whether the sublimated vapor in the capture tent will separate significantly from its attached regolith when it enters the liquefaction chamber. However, even if significant water-regolith separation does occur, distillation of the water-regolith solution will be required. Boiling with heated surfaces, vacuum distillation, flash evaporation, vapor compression, and direct and indirect heat transfer are all viable options for performing distillation in any environment, independent of its gravity. The most feasible distillation technique for moonbase water remains an open question.

### 6.1 Water Recovery Considerations

Among the human use activities from which wastewater is generated are food preparation, hygiene, and humidity condensate (produced by perspiration and respiration). Average consumable and wastewater values from two studies are adjusted for the 100-person moonbase and are shown in Table 1. Incongruities between consumed water and wastewater are due to the presence of water in consumed foods.

Table 1: Water use and daily waste for a 100-person moonbase. Adapted from [58].

| Event | Consumed Water (kg) [59] | Consumed Water (kg) [60] | Wastewater (kg) [59] |
|---|---|---|---|
| Drinking | 200 | 259 | - |
| Hygiene | 495 | 117 | 495 |
| Shower | 272 | 108 | 272 |
| Food Preparation | - | 103 | - |
| Operations | - | 16 | - |
| Dishwash | - | 354 | - |
| Clothes Wash | - | 195 | - |
| Humidity Condensate | - | - | 227 |
| Urine | - | - | 150 |
| Total | 967 | 1,152 | 1,144 |

In controlled studies, 90-97% water recovery has been shown to be possible in primary stage water recovery systems [61]. Unfortunately, increasing water recovery promotes biofouling and scaling. Consequently, primary processors will typically terminate water recovery when the solubility limit of wastewater solutions has been reached [61]. In spite of the biofouling and scaling problem, additional water recovery is made possible by brine dewatering technologies. This may present an additional challenge on a moonbase, however, because these technologies have greater energy per unit water volume requirements and need additional consumables beyond the primary recovery infrastructure.

However, there is a successful precedent that may negate the need for brine dewatering technologies – the Vapor Compression Distillation system onboard the International Space Station recovers in excess of 85% of wastewater while discarding the brine. If brine can be disposed of entirely, this is advantageous because the evaporative surfaces and wastewater lines can be corroded by concentrated brine as well as the pretreatment chemicals used to oxidate urine [61].

It is important to reiterate that the 90-97% water recovery estimates apply to human wastewater and do not consider water that is recoverable from the hydroponic operations. Studies have indicated that between 35.9% [62] and 67.12% [63] of hydroponic water can be recovered by rainwater harvesting and exhaust air water recovery, respectively. Clearly, rainwater recovery is not applicable to a moonbase, and as such, an estimate approaching the latter result of 67.12% is used here. In a scenario in which 95% of the human wastewater and 65% of the hydroponic transpiration water were recovered, the moonbase's water requirements would be reduced from 86.3 $acre \cdot feet \cdot yr^{-1}$ to 27.8 $acre \cdot feet \cdot yr^{-1}$. Consequently, the number of 10 kW heat drills decreases from 20,240 to 6,524, and the $170 \times 170\ m^2$ regolith sublimation field shrinks to $97 \times 97\ m^2$. Therefore, water recovery brings about a 67% reduction in both the number of heat drills and the area of the regolith sublimation field.

### 6.1.1 Urine Treatment

Human urine production ranges from 1-4 $L \cdot day^{-1}$ [64] and is composed of 95% water and 2.5% urea. Sodium, sodium chloride, creatinine, potassium, sulfur, ammonia, phosphorus, uric acid, and hippuric acid are present in smaller quantities [65]. Successful distillation of urine requires that the distillation temperature remain below ~340 K (150° F), otherwise the urea will be thermally decomposed [65]. Additionally, chemical pretreatment of the urine inhibits the formation of bacteria, and the associated fixing of free ammonia will help to maintain the sterility of the effluent water [65].

In the water treatment systems of the moonbase, it will be essential to chemically pretreat urine prior to it reaching the treatment processors. This could be done with potassium peroxymonosulfate[10] ($KHSO_5$) which is used onboard current NASA spacecraft. $KHSO_5$ lowers the urine pH to ~3 so as to avert bacterial propagation which would accelerate the formation of $NH_4^+$ and $NH_{3(g)}$ by hydrolyzing urea [66]. Urea hydrolysis has the additional problem of raising the urine pH which precipitates struvite[11] [67]. Furthermore, precipitation, crystal formation, and bacterial growth will lead to inescapable turbidity intensifications, further hindering the treatment of urine [68].

Fortunately, however, urine contains the plant-life-essential macro-elements – nitrogen, phosphorus, and potassium. Essential micro-elements including B, Mn, Mo, Cu, Fe, Zn, and Co are also present in urine. Therefore, many of the chemical elements which are vital to plant growth would be routinely excreted by the lunar colonists. However, it is noteworthy that the sodium concentration in urine is detrimental to plant growth. Untreated urine in a plant environment leads to water eutrophication, but when urine is used appropriately, a human-plant synergy will prosper both species [69].

---

[10] An off-white powder which has the textural appearance of white sugar.

[11] Struvite ($NH_4MgPO_4 \cdot 6H_2O$) is the most common mineral found in the urinary tract stones of humans, cats, and dogs.

## 6.2 A Concept for Sustainable Water Recovery

Current spacecraft water recovery technologies appear to be less applicable to a moonbase than do the corresponding terrestrial technologies. In the case of spacecraft life support systems, the minimizations of mass, energy, volume, and maintenance are paramount. These minimizations do not carry the same degree of essentiality for a moonbase. Furthermore, lunar gravity renders the non-trivial task of microgravity compatibility a moot point and may even be sufficient to aid in the water recovery process. A lunar colony will not have the ongoing ready access to Earth resupply missions that are available to the International Space Station. The moonbase will need to make do with locally available resources for water treatment. Fortunately, solar radiation and lunar regolith will be available in virtually unlimited quantities and for an endless duration. Therefore, a moonbase has water recovery requirements that are more akin to terrestrial environments than to low-Earth-orbit space stations.

Perka and Anderson have proposed a solar distillation system for a lunar colony [70]. However, this is a single component technology, and is therefore incomplete. Alternatively, Thomas et al. have proposed the 5-stage *Regolith Filter for Lunar Urine Solids in Habitats* (RFLUSH) water recovery system [58] (see Figure 12). In stage 1, wastewater (i.e., hygiene water, laundry water, food preparation water, urine, and humidity condensate) flows into the waste treatment systems. Stage 2 contains the wastewater flocculation tank where acclimated bacteria stimulate the hydrolyzation of urea resulting in ammonia reaction, biofilm growth, and mineral precipitation.

In stage 3, biomass and precipitates are sent through a granular filter using lunar regolith as its medium. Thomas et al. do not specify the degree to which external pressure would be required to operate their lunar regolith filter system, although they do acknowledge that RFLUSH’s stage 3 may not be entirely gravity operated. If lunar gravity alone is insufficient to keep stage 3 operating at the required flow rates, an external pressure module would need to be added. In stage 4, the filter effluent is disinfected with a solar distillation array. Finally, water postprocessing occurs in stage 5, where the water is polished and dispatched in potable form for use throughout the moonbase.

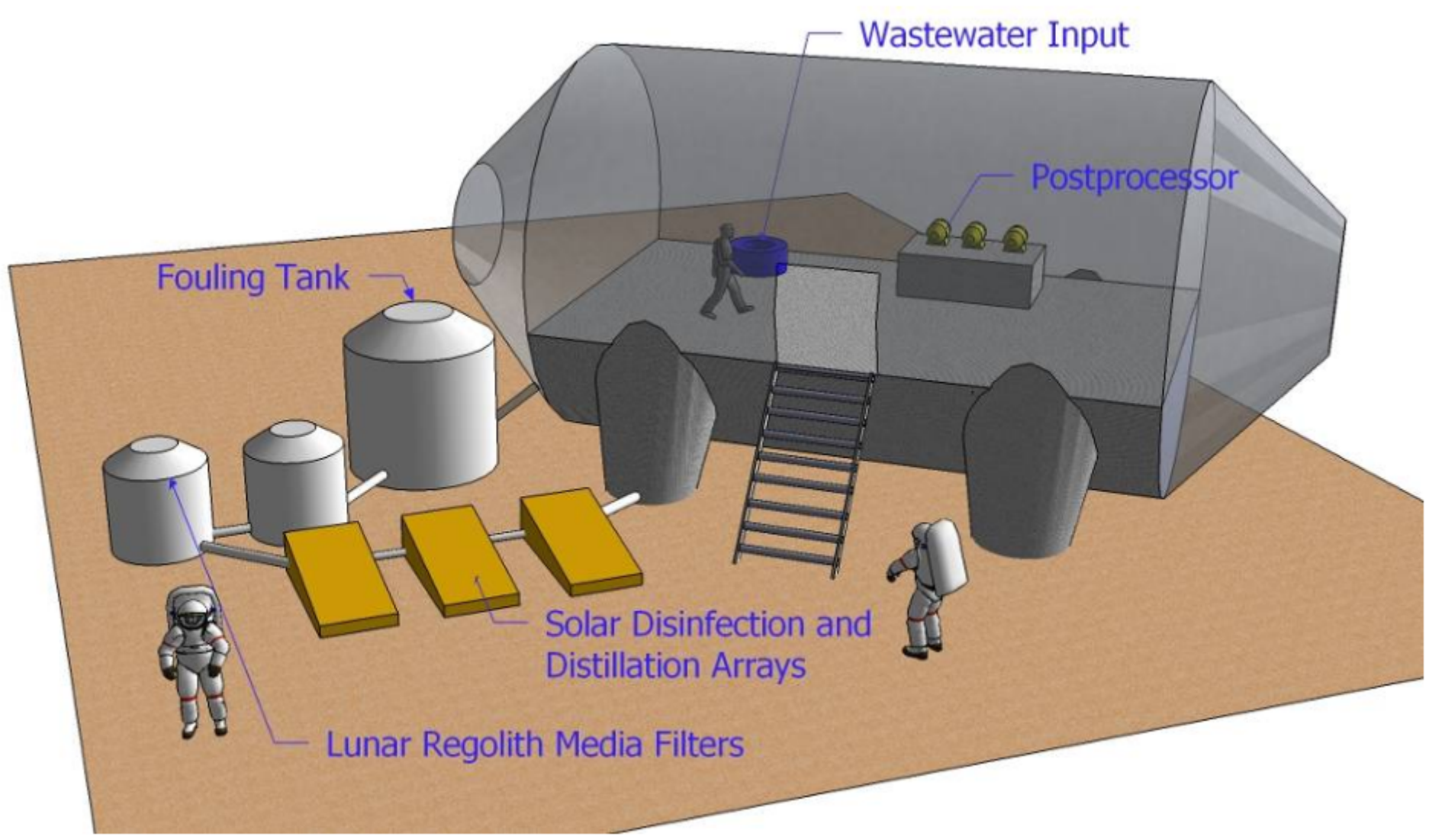

Figure 12: The RFLUSH water system of Thomas et al. [58].

Sustainability, use of in-situ resources, and integration are all essential requirements and are within the capability of the Regolith Filter for Lunar Urine Solids in Habitats. An important technology gap identified by NASA Exploration Life Support – “reduce resupply requirements by increasing loop closure and reducing consumable rates while increasing system reliability” [71] is well-addressed by RFLUSH.

## 6.3 Water Recovery Personnel

The staffing requirements for a municipal water treatment plant can provide a framework, but not a blueprint, upon which to build the staffing requirements for a lunar water treatment plant. In the lunar and terrestrial cases, labor productivity will depend predominantly on two factors: morale, which includes overall job satisfaction and pride in the plant operating efficiently; and job definitions/areas of work responsibility.

In terrestrial plants with effluxes $<10$ MGD, productivity will be highly dependent on morale. However, for terrestrial plants with effluxes $\geq 25$ MGD, eliciting a dependence of productivity on morale becomes problematic because of the highly complex staffing organization [72]. Even though the survival of the entire moonbase, including the treatment plant's personnel, is entirely contingent upon the uninterrupted access to potable water, innate human characteristics are location independent. Therefore, on a moonbase, the connection between productivity and morale is likely to remain largely intact.

The predominant difference in the backgrounds of the personnel employed at terrestrial water treatment facilities and their lunar counterparts is that every member of the latter group must first be astronauts. Only the very best in any given field are ever chosen to become astronauts. As such, the selected people who would service a moonbase's water treatment facility would be highly technically competent and stalwart team players. The rigorous screening, training, and selection processes employed by the world's space agencies do not eliminate the need for morale. Rather, the training is designed to teach methodologies for boosting morale. The "all in it together" mindset should be consistently reinforced because the survival of the lunar colony depends on it.

Whether the moonbase's organizational structure would be that of the military or just akin to the military is a pending decision that would be highly dependent upon the interplay between, and contributions of, the various civilian agencies, the military, private companies, and domestic and international consortiums involved in the moonbase's construction and operations. It appears inevitable that some kind of militaristic or semi- militaristic hierarchical configuration will be employed. Regardless of the underlying administrative structure, however, the overall responsibility for water production, treatment, and distribution would be that of the Water Resource Manager (WRM) who, as a member of the senior administration, would report directly to the Moonbase Commander.

Prior to assignment to the moonbase, eligible water treatment personnel would augment their general astronaut training with additional instruction that is specific to the lunar environment (including, but not limited to 1/6*g* training). Expertise and previous experience in terrestrial water management practices and the relevant technical services operations would also be required.

For a 1 MGD terrestrial trickling filter plant, the labor allocation is given in Table 2 and Table 3.

Table 2: Annual manhours of a 1 MGD terrestrial trickling filter plant [72].

| **Category** | **Annual Manhours** |
|---|---|
| Supervisory | 540 |
| Clerical | 50 |
| Laboratory | 350 |
| Yard | 470 |
| Operations | 1,690 |
| Maintenance | 1,600 |
| Total | 4,690 |

Table 3: Number of personnel required by a 1 MGD terrestrial trickling filter plant (assuming 1,500 manhours·yr$^{-1}$) [72].

| **Category** | **Number of Personnel** |
|---|---|
| Supervisory | 0.4 |
| Clerical | 0 |
| Laboratory | 0.2 |
| Yard | 0.3 |
| Operations | 1.1 |
| Maintenance | 1.1 |
| Total | 3.1 |

A proposed labor allocation for the above-mentioned personnel is [72]:

- Supervisor: 60% in operations and 40% in supervision.
- Operator: 50% in operations, 30% in maintenance, and 20% in lab work.
- Maintenance Personnel: 70% in maintenance and 30% in yard work.

Adapting the above figures to a moonbase is problematic. The moonbase water treatment facility will need to process only 24,820 GPD (27.8 acre·feet·yr$^{-1}$), but there are responsibilities which have no terrestrial analog. For example, the lunar heat drills will need to be relocated when the region they sublimate runs dry. It is likely that either a second capture tent will need to be erected, or the single capture tent will need to be relocated which would shut down the water treatment facilities. The latter option is not desirable.

The frequency of the heat drill relocation is uncertain, but the allotment of 15 additional personnel for heat drill maintenance and relocation, capture and liquefaction chamber servicing, and RFLUSH stage 3 regolith replenishment appears to be reasonable.

Therefore, approximately 20% of the complement are responsible for the acquisition, treatment, and management of the moonbase's water. This percentage will not scale with the size of moonbase's complement because an *n*-fold increase in moonbase personnel does not translate to an *n*-fold increase in the number of water production and treatment employees.

### 6.3.1 Water Management Responsibilities for "Operation Exodus"

In the event of a catastrophic failure of the moonbase's life support system, the moonbase may need to be completely evacuated. In this event, it would be the responsibility of the Moonbase Commander to initiate "Operation Exodus" – a total and potentially permanent evacuation of the moonbase. To facilitate Operation Exodus, the lunar colony would need to be equipped with approximately 10 "Ready Thirty"[12] emergency evacuation vehicles. It would be a responsibility of the water management personnel to make certain that all emergency escape shuttles remained continually stocked with potable water for a zero-notice return trip to Earth.

One example of a zero-notice evacuation catastrophe is the sudden and explosive decompression of a large region of the moonbase's structure. Such a catastrophe could be caused by the impact of a small asteroid. Although NASA's space shuttles were capable of surviving impacts that produced holes of up to 7 mm [73], the size of an asteroid impact that is survivable by the moonbase is uncertain.

However, were an unlikely but calamitous decompression to occur, the WRM would be one of a small group of senior administrators who would be required to make certain that all of their people immediately donned their spacesuits. If the Moonbase Commander ordered Operation

---

[12] Being ready to launch within 30 minutes.

Exodus, all senior administrators (including the WRM) would be responsible for ensuring that all of their people reached their assigned Ready Thirty escape shuttles forthwith.

In the event of a catastrophic failure of the water production and/or treatment facilities, the Water Resource Manager would be responsible for timely submitting his recommendation to the Moonbase Commander as to whether to implement Operation Exodus.

### 6.3.2 Moonbase Water Management: An Insight for Improving Terrestrial Water Management

Lunar water is a 4 Gyr-old legacy resource that makes possible mankind's first emigration from the Earth. The terrestrial atmosphere is capable of recycling water effectively, and as such, it makes water resources appear infinite with less apparent need for efficient management. However, water on Earth is not evenly distributed spatially or temporally; therefore, it requires significant management to meet the necessary quantity and quality in areas with growing populations and changing climates. The finitude of lunar water delivers a cautionary reminder that the water resources of our home planet are not infinite.

The examples of lunar water management necessary for a moonbase provide insight and inspiration for improving water management on Earth. The examples of efficient water management in food production with fewer trophic levels and more extreme recycling ideas on how water can be treated, recovered, and re-used could be modified and incorporated into more sustainable and efficient water use practices on Earth. In addition, the concept of centralized water management that appears to be an obvious necessity on the Moon may also be a helpful strategy in some situations on Earth.

## Conclusion

For a 100-person self-sustaining moonbase adjacent to the Shackleton-de Gerlach Ridge, the human, agricultural, and technical water requirements have been determined to be 12.3, 72, and 2 acre·feet·$yr^{-1}$, respectively. The Lunar Exploration Neutron Detector, the Diviner Radiometer, and the Lyman Alpha Mapping Project instrument onboard the Lunar Reconnaissance Orbiter have provided effectively irrefutable evidence of ~486,000 acre·feet of both surface and subsurface water ice at the lunar poles, with the majority of that ice being located at the lunar south pole.

A 3D numerical model has revealed that 6,524 10 kW heat drills quadratically distributed across a 97 × 97 $m^2$ regolith sublimation field can sublimate sufficient water ice to meet the needs of the lunar colony.

The sublimated materials were ejected from the regolith into a collection tent (which covers the regolith sublimation field), liquified in an adjacent liquefaction chamber, made potable by the 5-stage RFLUSH treatment technology, and distributed to the lunar colony.

Approximately 20 moonbase personnel, specially trained in technical service activities in lunar gravity, would oversee the sublimation and treatment facilities around the clock and report to the Water Resource Manager.